\documentclass[a4paper,11pt]{article}
\pdfoutput=1

\usepackage{amsmath}

\usepackage{jcappub}
\usepackage{bbold}

\usepackage{subfigure}
\usepackage{graphicx}
\usepackage{float}
\usepackage{xcolor}

\def\be{\begin{equation}}
\def\ee{\end{equation}}
\def\bea{\begin{eqnarray}}
\def\eea{\end{eqnarray}}
\def\ba{\begin{align}}
\def\ea{\end{align}}
\def\bi{\begin{itemize}}
\def\ei{\end{itemize}}

\def \al {\alpha}
\def \de {\delta}
\def \ka {\kappa}

\def \ga {\gamma}

\def \la {\lambda}

\def \Si {\Sigma}
\def \de {\delta}
\def \De {\Delta}

\def \Om {\Omega}
\def \dd {\partial}
\def \HH{\mathcal H}
\def \ud {\mathrm d}
\def \Perp {\mathcal P}

\newcommand{\bk}{{\bf k}}
\newcommand{\bn}{{\bf n}}

\newcommand{\thJ}[6]{\begin{pmatrix} #1&#2&#3 \\ #4&#5&#6 \end{pmatrix}}

\title{Higher order relativistic galaxy number counts: dominating terms}

\author[a]{Jeppe Tr\o st Nielsen}
\affiliation[a]{Niels Bohr International Academy, Blegdamsvej 17, Copenhagen 2100, Denmark}
\author[b]{and Ruth Durrer}
\affiliation[b]{University of Geneva, Department of Theoretical Physics and Center for Astroparticle Physics, 24 quai E. Ansermet, CH-1211 Geneva 4, Switzerland }

\emailAdd{Jeppe.Trost@nbi.dk}
\emailAdd{Ruth.Durrer@unige.ch}

\abstract{ We review the number counts to second order concentrating on the terms which dominate on sub horizon scales. We re-derive the result for these terms and compare it with the different versions found in the literature. We generalize our derivation to higher order terms, especially the third order number counts which are needed to compute the 1-loop contribution to the power spectrum.}

\begin{document}
\maketitle
\flushbottom

\section{Introduction}

About two years ago, nearly simultaneously, three different derivations of the second order number counts have been published~\cite{Bertacca:2014wga,Yoo:2014sfa,DiDio:2014lka,Bertacca:2014hwa}. Ref.~\cite{Bertacca:2014hwa} is a follow up of Ref.~\cite{Bertacca:2014wga}
completing several aspects of the previous work and in particular including magnification bias.

Even though these works are potentially interesting, they are very hard to compare as the final formula occupies several pages and, especially in Ref.~\cite{Yoo:2014sfa}, has to be pieced together from a multitude different contributions. Also the notation and the break-up into separate terms is very different in all the derivations. An additional difficulty is that terms can be converted into each other by integrations by part, often in a quite non-trivial way, rendering the comparison of partial results tricky. 

Because of these difficulties, in this paper we do not attempt to compare the full formulas but we concentrate on the terms which dominate on sub horizon scales. As proposed in Ref.~\cite{DiDio:2015bua} we only consider terms of the order of $(k/\HH)^4\Psi^2$ and neglect smaller contributions to the second order number count. Here $k$ is the comoving wave number, $\HH$ the conformal Hubble parameter and $\Psi$ the Newtonian potential or more precisely the Bardeen potential.
As we shall show in this paper, these terms can be identified  uniquely and they have a simple physical interpretation.

Using a rather straight forward  derivation we reproduce the result of Ref.~\cite{DiDio:2014lka}. We also compare the corresponding terms in the two other derivations and we find some disagreement with both Refs.~\cite{Bertacca:2014wga} and \cite{Yoo:2014sfa}. The disagreements concern lensing terms and a double counting of volume distortion effects, respectively. The authors of these refs. agree with our findings and in the latest arXiv version of  Ref.~\cite{Bertacca:2014wga} the error is corrected.

Furthermore, the simple interpretation of the dominant terms allows us to derive the form of higher order dominant corrections to the number counts. We partially check this result comparing with the third order computation of  angular perturbations from Ref.~\cite{Fanizza:2015swa}, with which it agrees.

The remainder of this paper is organized as follows. In the next section we derive a rather simple formula for the dominant contributions to second order number counts and compare it with the above mentioned literature. In Section~\ref{sec:higherorder} we generalize this formula to higher orders. In Section~\ref{s:num} we present a preliminary study of one loop  contributions to the power spectrum and in Section~\ref{s:con} we conclude.

{\bf Notation: ~} We set $c=1$. We consider a flat Universe with only scalar perturbations, so that the metric is given by
\be
ds^2 =a^2\left[-(1+2\Psi)dt^2 +(1-2\Phi)\de_{ij}dx^idx^j\right] \,.
\ee
We set $\Phi=\Psi$, neglecting anisotropic stresses as their contribution is subdominant. Also the vector and tensor perturbations induced at second order are subdominant in our counting and will not be discussed here. We split all perturbative quantities as 
$$
F = \sum_n F^{(n)},
$$
with no factor $\frac{1}{n!}$.
The notation keeping perturbative orders may become cumbersome. However we deem it necessary to keep explicit track of it, in particular when going to third order, where the order of individual terms is not obvious.

\section{Dominant terms in the second order number counts: A comparison of the result in the literature}\label{s:deriv}

\subsection{Derivation of the main result of \citet{DiDio:2014lka}}
The dominant contributions to the first order number counts have originally been derived in~\cite{Matsubara:2000pr}. The full relativistic expressions including all terms can be found in~\cite{Yoo:2009au,Bonvin:2011bg,Challinor:2011bk}.
Here we use the formula by~\cite{Matsubara:2000pr} which gives
\be\label{e:1stnumber}
\Si^{(1)}(\bn,z) = b\de^{(1)}(r(z)\bn,t(z)) + \HH^{-1}\dd_r^2v^{(1)}(r(z)\bn,t(z)) -2\ka^{(1)}(r(z)\bn,t(z)) \,.
\ee 

$\Si^{(1)}$ is first order number count fluctuation in direction $\bn$ at observed redshift $z$, $\de$ is the matter density fluctuations and $b$ the bias factor which may be redshift dependent, $r(z)$ denotes the comoving distance out to redshift $z$ and $t(z)$ denotes conformal time at redshift $z$.  The full first order number count perturbation is of the form
\be
\De^{(1)}(\bn,z) =\Si^{(1)}(\bn,z) +{\rm LS},
\ee
where '$\rm LS$' contains terms of order $(k/\HH)^n\Psi^{(1)}$ with $0\le n<2$  which are observable only on very large scales, $k\sim \HH(z)$ and which we neglect in our discussion.
Their observation is an interesting topic which is presently being addressed in the literature~\cite{Camera:2015yqa,Alonso:2015uua,Alonso:2015sfa,Irsic:2015nla}.
The second term in Eq.~(\ref{e:1stnumber}) is the well known redshift space distortion (RSD)~\cite{Kaiser:1987qv}, where $v$ is the velocity potential and $-\dd_rv$ the radial velocity. It represents the radial volume distortion. The third term, finally comes from lensing which affects the area subtended by a given solid angle. It is the transversal volume distortion. Denoting the  direction  of observation at the observer by $\bn =(\theta_0^1,\theta_0^2)$ and the direction at the source position $z$ by $(\theta^1_z,\theta^2_z)$, the lens map maps $(\theta_0^1,\theta_0^2) \mapsto (\theta_z^1,\theta_z^2)$. For scalar perturbations its Jacobian is given to linear order by~\cite{Bartelmann:1999yn}
\be\label{e:Aab}
{A^a}_{b} \equiv\frac{\dd\theta_z^a}{\dd\theta_0^b} \simeq {\de^a}_{b}-\nabla_b\nabla^a\phi^{(1)} \,,
\ee
where $\phi^{(1)}$ is the first order lensing potential. We introduce the $n$th order lensing potential by
\be\label{e:lenspot}
\phi^{(n)}(\bn,z) =-2\int_0^{r(z)}dr\frac{r(z)-r}{r(z)r}\Psi^{(n)}(r\bn,t(z)) \,,
\ee
and $\nabla_a$ is the covariant derivative in direction $\theta^a$ on the sphere.
$A$ can be split into its trace and its traceless part which for scalar perturbation is of the form
\be
{A^a}_{b} = \left(\begin{array}{cc}  1-\ka-\ga_1 & \ga_2\\ \ga_2 & 1-\ka+\ga_1 \end{array} \right)\,.
\ee
The area subtended by a given solid angle is the determinant of $A$ given by
\be
| A | = (1-\ka)^2 -|\ga|^2 \simeq 1-2\ka^{(1)} \,.
\ee
Here the expression after the $\simeq$ sign is the first order approximation which is taken into account in Eq.~(\ref{e:1stnumber}) and $\ga = \ga_1+i\ga_2$ is the complex shear such that $|\ga|^2 = \ga_1^2 +\ga_2^2$. The convergence $\ka$ is 
\be
\ka^{(1)}(\bn,z) = - \frac{1}{2}\nabla^a\nabla_a\phi^{(1)} = \int_0^{r(z)}dr\frac{r(z)-r}{r(z)r}\De_\Om\Psi^{(1)}(r\bn,t(z))\,,
\ee
where $\De_\Om$ denotes the Lapacian on the sphere.

Inspecting the second order relativistic perturbation equations derived in Ref.~\cite{Acquaviva:2002ud}, it is easy to see that the dominant term in the second order gravitational potential is simply the one coming from the second order density fluctuations such that
\be
k^2\Psi = -4\pi G\bar\rho a^2 (\de^{(1)} + \de^{(2)}) = -\frac{3}{2}\HH^2(z)\Om_m(z) (\de^{(1)} + \de^{(2)}) = k^2(\Psi^{(1)}+\Psi^{(2)}) \,.
\ee
Here $\de^{(1)}$ is the first order density fluctuation and  $\de^{(2)}$ is the Newtonian second order density fluctuation.
The Newtonian second order density and velocity perturbation are given e.g. in Ref.~\cite{Bernardeau:2001qr}. We present them in Appendix~\ref{app:Newton} for completeness. 

Like for the first order, the expression for $\Sigma(\bn,z)$ to second order is split up as
\be  \De^{(2)}(\bn,z)=  \Sigma^{(2)}(\bn,z)+{\rm LS},
\ee
where '$\rm LS$' contains terms of order $(k/\HH)^n\Psi^{(1)2}$ with $0\le n<4$  which are relevant only on very large scales, $k\sim \HH(z)$ and which we neglect in our discussion. To determine $ \Sigma^{(2)}$
we must consider the following:
\begin{enumerate}
\item The general formula for the dominating terms is of the following form
\begin{equation}
\Sigma = (1+\delta)(1+\de V) = (1+\delta)(1+\text{\textsc{rsd}}) |{A^a}_b| \, .
\end{equation}
\item The second order contribution is
\begin{equation}
\Sigma^{(2)} = [\delta]^{(2)}+[\text{\textsc{rsd}}]^{(2)} + [|{A^a}_b|]^{(2)} + \delta^{(1)}\text{\textsc{rsd}}^{(1)} +\delta^{(1)} |{A^a}_b|^{(1)}+\text{\textsc{rsd}}^{(1)} |{A^a}_b|^{(1)}.
\end{equation}
\item Second order components in this formula must be treated consistently. This in particular includes a Taylor expansion of first order perturbations to account for the first order shifts in the radial and transversal position. Non-integrated quantities $F(r(z)\bn,t(z))$ to second order at perturbed coordinates is therefore
\begin{align}\label{e:2ndexp}
[&F( r(z+\de z)(\bn+\de\bn) )]^{(2)} \simeq 
 \left[F^{(2)}+ \HH^{-1} \dd_r v^{(1)} \dd_{r} F^{(1)}+ \nabla^a\phi^{(1)}\nabla_a F^{(1)}\right](r(z)\bn), 
\end{align}
where we have inserted the transversal shifts $\theta^{(1)a} = \nabla^a\phi^{(1)}$,  and the radial shift $r^{(1)} = H^{-1} z^{(1)} = \HH^{-1} \dd_r v^{(1)}$. In our notation $z$ is the observed redshift while the unperturbed redshift relevant for computing the comoving distance $r$ is $z+\de z$. Similarly, $\bn$ is the observed radial direction while $\bn+\de\bn$ is the true direction of the source. We have neglected time derivatives, which are subdominant in our counting, spatial derivatives typically generate a factor $k$ while time derivatives scale like $\HH$.

\item For a function integrated along the line of sight, we must expand the position inside the integral, meaning we go beyond the Born approximation and expand the deviation of the photon path to first order, i.e., we integrate $f(\bn r)$ \emph{along the perturbed trajectory},
\bea
\left[\int_0^{r(z+\de z)}dr f((\bn+\de\bn) r)\right]^{(2)} &\simeq& \int_0^{r(z)}\!drf^{(2)} +  \int_0^{r(z)}\!dr\de\theta^a\nabla_af^{(1)}(\bn r) \nonumber \\
&=& \int_0^{r(z)}\!drf^{(2)} +  \int_0^{r(z)}\!dr\nabla^a\phi^{(1)}\nabla_af^{(1)} \,.
\eea
We discard the radial distortions of the integral. It is subdominant as in the integral along the line of sight we can trade a radial derivative with a boundary term and a time derivative which both do not have a factor $k$ and are therefore subdominant. Hence for all integrated terms like the deflection angle or the lensing potential we can neglect radial derivatives.
\end{enumerate}
This gives a clear separation of the leading terms. All the first order cross-terms are readily written down
\begin{equation}
\delta^{(1)}\text{\textsc{rsd}}^{(1)} +\delta^{(1)} |{A^a}_b|^{(1)}+\text{\textsc{rsd}}^{(1)} |{A^a}_b|^{(1)} = 
\HH^{-1}\delta^{(1)}\dd_r^2 v^{(1)} -2\delta^{(1)} \kappa^{(1)} -2\HH^{-1}\dd_r^2 v^{(1)}\kappa^{(1)} \, .
\end{equation}
The second order density contrast is
\begin{equation}
[\de(\bn+\de\bn,z+\de z)]^{(2)} = \de^{(2)} + \HH^{-1} \dd_r v^{(1)} \dd_{r} \de^{(1)} + \nabla^a\phi^{(1)}\nabla_a \de^{(1)},
\end{equation}
where we drop the argument $(\bn,z)$ on the right hand side. To compute the second order RSD we must remember that  it appears as the derivative of the redshift-perturbation (see Ref.~\cite{Bonvin:2011bg}, Eq.~(17)), and we must expand \emph{that} to compute the second order:
\begin{align}
&\text{\textsc{rsd}}^{(1)} = H^{-1} \dd_r z^{(1)} \Rightarrow\nonumber \\
[&\text{\textsc{rsd}}(\bn+\de\bn,z+\de z)]^{(2)} = \HH^{-1}\dd_r
\left( \dd_r v^{(2)} + \HH^{-1} \dd_r v^{(1)} \dd_r^2 v^{(1)} + \nabla^a\phi^{(1)}\nabla_a \dd_r v^{(1)} \right) \, .
\end{align}
A short calculation shows that the second order lensing term can be written as
\begin{align}\label{e:lensref}
[|{A^a}_b|]^{(2)} = \nabla_a[\theta^a]^{(2)} + \kappa^{(1)2} -|\ga^{(1)}|^2 = \nabla_a[\theta^a]^{(2)}  + 2\kappa^{(1)2} - (\nabla_b\nabla_a\phi^{(1)})(\nabla^b\nabla^a\phi^{(1)})/2 \, .
\end{align}
The first term is computed by expanding the angular derivative inside the integral,
\begin{align}
\nabla_a\theta^{(1)a} &= -2 \nabla_a \int_0^{r(z)}dr\frac{r(z)-r}{r(z)r}\nabla^a\Psi^{(1)} \Rightarrow \nonumber \\
\nabla_a[\theta^a(\bn+\de\bn,z+\de z)]^{(2)} &= -2\nabla_a \int_0^{r(z)}dr\frac{r(z)-r}{r(z)r}\left( \nabla^a\Psi^{(2)} + \nabla^b\phi^{(1)}\nabla_b \nabla^a\Psi^{(1)} \right)  +{\rm LS}\nonumber \\
 &=\De_\Omega \phi^{(2)} -2 \int_0^{r(z)}dr\frac{r(z)-r}{r(z)r}\nabla_a\left( \nabla^b\phi^{(1)}\nabla_b \nabla^a\Psi^{(1)} \right) + {\rm LS} \nonumber\\
&= -2\kappa^{(2)} -2 \int_0^{r(z)}dr\frac{r(z)-r}{r(z)r}\nabla_a\left( \nabla^b\phi^{(1)}\nabla_b \nabla^a\Psi^{(1)} \right) + {\rm LS} \, .\label{e:angle2} 
\end{align}
This accounts for the leading second order terms.

Neglecting the sub-leading terms and rewriting the remaining, with some integrations by part in the pure lensing term given in the last two lines below, we find
\begin{align}\label{e:reference}
\Sigma^{(2)} =&\  \HH^{-1}\delta^{(1)}\dd_r^2 v^{(1)} -2\delta^{(1)} \ka^{(1)} -2\HH^{-1}\dd_r^2 v^{(1)}\ka^{(1)}  \nonumber\\
&\ +  \de^{(2)} + \HH^{-1} \dd_r v^{(1)} \dd_{r} \de^{(1)} + \nabla^a\phi^{(1)}\nabla_a \de^{(1)} \nonumber \\
		&\  +\HH^{-1} \dd_r^2 v^{(2)} + \HH^{-2}\dd_r \left( \dd_r v^{(1)} \dd_r^2 v^{(1)}\right) + \HH^{-1}\nabla^a\phi^{(1)}\nabla_a \dd_r^2 v^{(1)} \nonumber \\
&-2  \ka^{(2)} + 2 \ka^{(1)2} 
-2\nabla_a \ka^{(1)} \nabla^a \phi^{(1)}  \nonumber \\ &\ 
 - \frac{1}{2r(z)} \int_0^{r(z)} \!\!dr\frac{r(z)-r}{r} \Delta_\Om \left( \nabla^a \Psi_1^{(1)} \nabla_a \Psi_1^{(1)} \right)
-2 \int_0^{r(z)} \frac{dr}{ r}  \nabla^a \Psi_1^{(1)}\nabla_a \ka^{(1)} \,,		
\end{align}
where we have introduced $\Psi_1 \neq \Psi$ given by
\begin{equation}
\Psi_1^{(n)} = -r \frac{d\phi^{(n)}}{dr} \, .
\end{equation}
This matches exactly the results of Refs.~\cite{DiDio:2014lka, DiDio:2015bua}.

Let us comment on the way that this result is affected by biasing. In the real Universe we do not observe density fluctuations but galaxies. The relation between the galaxy over density and the matter density is what we call 'biasing'. Galaxy bias is in general not linear, but if we assume it to be local it can be cast in the following ansatz~\cite{Bernardeau:2001qr} for the galaxy over density $\de_g$
\be
\de_g = \sum_k \frac{b_k}{k!}\de^k \,.
\ee
Up to second order this yields
$$
\de_g^{(1)}+\de_g^{(2)} = b_1\de^{(1)} + b_1\de^{(2)} + \frac{b_2}{2}\left(\de^{(1)}\right)^2 \,.
$$
In principle, in full General Relativity, a discussion of the correct gauge for the biasing scheme has to follow here. But since the corrections coming from gauge transformations are subdominant and neglected in our counting, we can ignore this difficulty. 
When expressing the number counts for galaxies we have therefore to replace $\de^{(2)}$ by $\de_g^{(2)} =  b_1\de^{(2)} + \frac{b_2}{2}\left(\de^{(1)}\right)^2$ and $\de^{(1)}$ by $\de_g^{(1)} =  b_1\de^{(1)}$ everywhere. This modifies (\ref{e:reference}) to
\begin{align}\label{e:reference-bias}
\Sigma_g^{(2)} =&\  \HH^{-1}b_1\delta^{(1)}\dd_r^2 v^{(1)} -2b_1\delta^{(1)} \ka^{(1)} -2\HH^{-1}\dd_r^2 v^{(1)}\ka^{(1)}  \nonumber\\
&\ +  b_1\de^{(2)} +\frac{b_2}{2} \left(\de^{(1)}\right)^2+ \HH^{-1} b_1\dd_r v^{(1)} \dd_{r} \de^{(1)} + b_1\nabla^a\phi^{(1)}\nabla_a \de^{(1)} \nonumber \\
		&\  +\HH^{-1} \dd_r^2 v^{(2)} + \HH^{-2}\dd_r \left( \dd_r v^{(1)} \dd_r^2 v^{(1)}\right) + \HH^{-1}\nabla^a\phi^{(1)}\nabla_a \dd_r^2 v^{(1)} \nonumber \\
&-2  \ka^{(2)} + 2 \ka^{(1)2} 
-2\nabla_a \ka^{(1)} \nabla^a \phi^{(1)}  \nonumber \\ &\ 
 - \frac{1}{2r(z)} \int_0^{r(z)} \!\!dr\frac{r(z)-r}{r} \Delta_\Om \left( \nabla^a \Psi_1^{(1)} \nabla_a \Psi_1^{(1)} \right)
-2 \int_0^{r(z)} \frac{dr}{ r}  \nabla^a \Psi_1^{(1)}\nabla_a \ka^{(1)} \,.		
\end{align}
Here we explicitly only include galaxy bias and not magnification bias. We also neglect velocity bias.
Note the addition of the positive definite term $b_2 \left(\de^{(1)}\right)^2/2$. In Fourier space it is this term which renders $\langle\de^{(2)}_g\rangle \neq 0$, as the mean of the intrinsic terms vanishes in Fourier space, $\langle \de^{(2)}\rangle = \langle v^{(2)}\rangle =\langle \ka^{(2)}\rangle =0$. This is not the case for $\Si^{(2)}$. Even if we neglect second order bias, we obtain after several integrations by part, that the angular average of $\Si^{(2)}$ does not vanish. Setting $b_1=1$ we obtain
\be\label{e:average}
\left\langle\Si^{(2)}\right\rangle = \left\langle\HH^{-1} \dd_r ( \delta \dd_r v) + \HH^{-2} \dd_r ( \dd_r v \dd_r^2 v ) -4 \kappa^2\right\rangle
\ee
A short calculation in Fourier space shows that the
 two first terms vanish (this is simply a consequence of statistical isotropy), and we are left with the lensing term $\langle-4\kappa^{(1)2}\rangle$.  
 The sign of this term can be understood as follows: convergence increases the angle under which we see the images of two galaxies. This leads us to infer a larger transverse volume and hence a lower density.
 
 For a constant bias, the lensing term is in general much smaller than the bias-induced term $\left\langle \de^{(1)2}/2\right\rangle$. However, if $b_2(z)\propto 1/D(z)^2$, as a conserved number of galaxies would imply, the lensing term dominates at high redshift. 
 
 Note that even though in $k$-space a zero mode is simply subtracted and one considers $\de^2-\langle\de^2\rangle$ which by definition has no zero-mode.
 In redshift space this is not so simple. A zero-mode like $\langle\de^2\rangle$ typically depends on redshift and the quantity $C_0(z,z')$ which contains the zero-mode does contain physical information, see e.g.~\cite{DiDio:2013sea} for a discussion. It is therefore useful not to perform the redshift dependent subtraction of $\langle\de^2\rangle$. This is even more true for $\kappa^2$ which has a non-trivial redshift dependence.
 
 \begin{figure}[htb]
\begin{center}
\includegraphics[width=.6\textwidth]{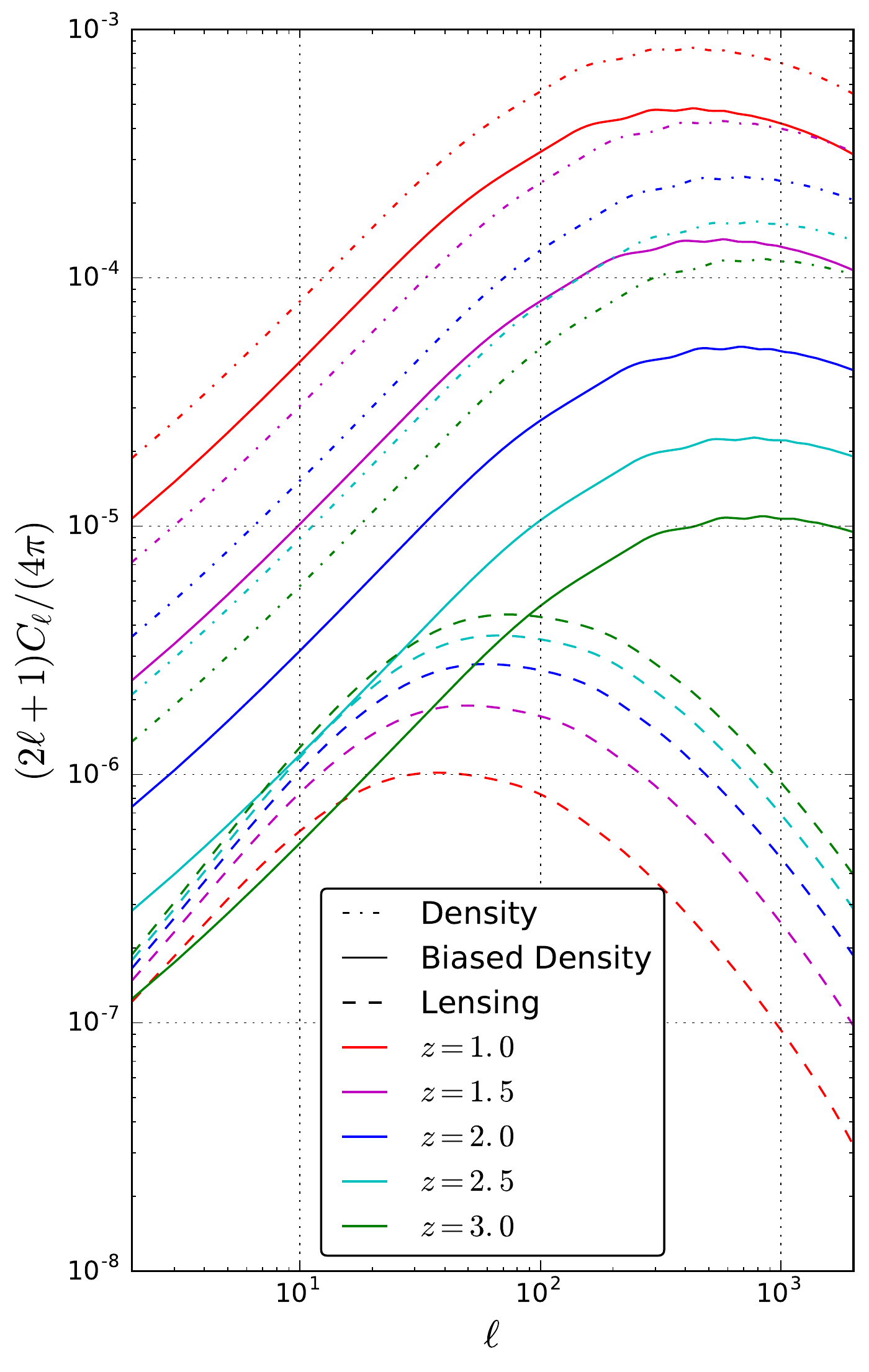}
\end{center}
\caption{ \label{fig:denslens} Log-log plot of the first order $C_\ell$'s the sum of which yields the biased density-density terms. We compare two cases, $b_2=$constant (dash-dotted lines) and $b_2\propto D^{-2}(z)$ which is requested for a conserved number of galaxies (full lines)  with $b_2$ normalized to unity today.  We also plot the  first order power spectrum of the lensing term $4\kappa^{(1)2}$ (dashed lines). The   second order one-point function is given by the sum of $(2\ell+1)C^\kappa_\ell/(4\pi)$ over $\ell$. We see that as the redshift increases, the density contribution decreases. The lensing contribution on the other hand is enhances with growing redshift. While the sum for the lensing term nicely converges, the second order bias term requires a counter-term for convergence. (The numerical results have been obtained with {\sc class}~\cite{DiDio:2013bqa},  the Limber approximation is used for the $\ka$-spectrum but not for $\de$.)} 
\end{figure}
In Fig.~\ref{fig:denslens} We show the first order angular spectra of $\kappa$ and $\delta$ for different redshifts. The zero modes discussed above are simply given by
$$b_2\langle \delta^2\rangle = \sum _\ell(2\ell+1)b_2C_\ell^{\delta\delta}\,,\quad 
 4\langle \kappa^2\rangle =\sum _\ell(2\ell+1)4C_\ell^{\kappa\kappa}\,. $$
  For a second order bias of $b_2\sim 1$, the bias induced term always dominates. However, the lensing term grows with redshift while the second order bias term decays.

\subsection{Comparison with~\citet{Bertacca:2014wga}}
In the following, we identify and match terms from Ref.~\cite{Bertacca:2014wga} to Eq.~\eqref{e:reference}. We start by relating the notation of one to the other. The main notational differences are:
\begin{itemize}
\item The second order metric perturbations are defined with a factor 2 difference.
\item There is a factor $(-1)$ between the definitions of the velocity potential.
\item Projected derivatives are used, ie. $\dd_\| = \dd_r$ and $\dd^i_\perp = r^{-1}\nabla^a$.\footnote{Note that the index structure here enjoys a slight abuse of notation. However, the angular derivatives always appear contracted, and we \emph{do} have the strict equality $\dd^i_\perp f \dd_{i\perp} g =  r^{-2}\nabla^a f \nabla_a g$.}
\end{itemize}

Identifying the leading terms in the expression for $\De^{(2)}$, and expanding the expression for $\De^{(1)}$, we find 
\begin{align}\label{e:transSigma2}
\Sigma^{(2)} =&\ \delta_g^{(2)} -\frac{1}{ \HH }\dd_\|^2v^{(2)}   - 2\kappa^{(2)} +4\kappa^{(1)2} -4\delta\kappa^{(1)} - 2\frac{\delta_g}{\HH} \dd_\|^2 v \nonumber \\
&+ \frac{4\kappa^{(1)}}{\HH}\dd_\|^2 v + \frac{2}{\HH^2}\left(\dd_\|^2 v  \right)^2 + \frac{2}{\HH^2}\dd_\| v \dd_\|^3v  - \frac{2}{\HH}\frac{\ud \delta_g }{\ud  \bar \chi} \Delta \ln a^{(1)} \nonumber \\
& -2\bigg[ \bar \chi  \dd_{\perp i}\delta_g -\frac{\bar \chi}{\HH}\dd_{\perp i} \dd_\|^2 v \bigg]  \dd_\perp^i T^{(1)}
+4 \bigg[ \bar \chi  \dd_{\perp i}\delta_g -\frac{\bar \chi}{\HH}\dd_{\perp i} \dd_\|^2 v \bigg]  S^{i(1)}_\perp \nonumber \\
& -4 \bigg(\int_0^{\bar \chi} \ud \tilde\chi \frac{\tilde \chi}{ \bar \chi}
 \left(\bar \chi-\tilde \chi\right) \Perp^p_j \Perp^{iq} \tilde \dd_q  \tilde \dd_p\Phi \bigg)
  \bigg(\int_0^{\bar \chi} \ud \tilde\chi \frac{\tilde \chi}{ \bar \chi}
 \left(\bar \chi-\tilde \chi\right) \Perp^n_i \Perp^{jm} \tilde \dd_m  \tilde \dd_n\Phi \bigg) \, .
  \end{align}
The translation of the majority of the terms is straight-forward.  $a^{(1)}$ is the first order perturbation of the scale factor taken to be $1/(z+1)$. To leading order, we may substitute it by the redshift space distortion, $\De\ln a^{(1)} = -\dd_r v^{(1)}$. $\bar\chi$ is the comoving distance which we call $r$ and  $\ud/\ud\bar\chi=-\ud/\ud\la=\dd_r -\dd_t\simeq \dd_\|$, the difference $\dd_t$ is subdominant in our counting. With this we can write the first two lines immediately in the more familiar form,
\begin{align}
& \delta^{(2)} + \frac{1}{ \HH }\dd_r^2v^{(2)} - 2\kappa^{(2)} +2\bigg(2\kappa^{(1)2} -2\delta\kappa^{(1)} + \frac{\delta^{(1)}}{\HH} \dd_r^2 v^{(1)} \nonumber \\
&- \frac{2\kappa^{(1)}}{\HH}\dd_r^2 v^{(1)} + \frac{1}{\HH^2}\left(\dd_r^2 v^{(1)}  \right)^2 + \frac{1}{\HH^2}\dd_r v^{(1)} \dd_r^3v^{(1)}  + \frac{1}{\HH} \dd_r \de^{(1)} \dd_r v^{(1)} \bigg) \, .
\end{align}
The third line above requires the translation of $S^{i(1)}_\perp$ and $\dd_\perp^i T^{(1)}$.  Denoting the transverse direction  of a vector, $v^i_\perp$ by $v^a$, these are 
\begin{align}
S^{i(1)}_\perp &= -\nabla^a\int^{r_s}_0 \ud r \frac{1}{r} \Psi^{(1)} \, ,\qquad
\dd_\perp^i T^{(1)} = - \nabla^a\int^{r_s}_0 \ud r \frac{2}{r_s} \Psi^{(1)} \nonumber \\
&\mbox{so that} \qquad 2 S^{i(1)}_\perp - \dd_\perp^i T^{(1)} = \nabla^a \phi^{(1)} \, .
\end{align}
This multiplies the term in brackets, hence this line is the angular Taylor expansion of $\delta + \HH^{-1}\dd_r^2 v$, with the appropriate extra factor $2$. In the last line of \eqref{e:transSigma2} we simply need to substitute the derivatives and notation of potentials to obtain
\begin{equation}
 -  \nabla^a  \nabla^b\bigg(2\int_0^{\bar \chi} \ud \tilde\chi \frac{\bar \chi-\tilde \chi}{ \bar \chi \tilde\chi}
 \Psi^{(1)} \bigg)
 \nabla_a  \nabla_b \bigg(2\int_0^{\bar \chi} \ud \tilde\chi \frac{\bar \chi-\tilde \chi}{ \bar \chi \tilde\chi}
  \Psi^{(1)} \bigg)
  = -2 \frac{1}{2} \nabla^a\nabla^b \phi^{(1)} \nabla_a\nabla_b\phi^{(1)} \, .
\end{equation}
This is exactly twice the last term of Eq.~\eqref{e:lensref}, which cancels a term in $\nabla_a[\theta^a]^{(2)}$, and therefore does not appear in the final result. The factor $2$ comes from the definition of the second order perturbations. 

This calculation extends the footnote 1 of Ref.~\cite{DiDio:2015bua}, and shows that the difference between the results of Refs.~\cite{DiDio:2015bua} and \cite{Bertacca:2014wga} is simply the substitution of $\nabla_a[\theta^a]^{(2)}$ with $\kappa^{(2)}$, which neglects the terms coming from the fact that in $\nabla_a[\theta^a]^{(2)}$ there are the additional lensing terms, namely all terms coming from evaluating the first order lensing integral at the perturbed position leading to the second  term of Eq.~\eqref{e:angle2}, the so called 'post-Born' contributions\footnote{The authors of Ref.~\cite{Bertacca:2014wga} agree with this finding and they have corrected the error in v4 on the arXiv with which we fully agree.}.

\subsection{Comparison with~\citet{Yoo:2014sfa}}
We now proceed to identify the leading terms of Ref.~\cite{Yoo:2014sfa}. Since the full final result is not written down in closed form, we perform the same 'stitching-together' as in the beginning of  Sec.~\ref{s:deriv}, working our way back through the paper. The main remarks about the notation here are:
\begin{itemize}
\item Latin indices go from 0 to 3, greek indices go from 1 to 3.
\item The metric perturbations are called $\mathcal A$ and $\mathcal C_{\alpha\beta}$, where ${\mathcal C}_{\alpha\beta} = -\Psi\de_{\alpha\beta}$ in the gauge we are working, for purely scalar perturbations. They are however defined with no factor $2$ at second order. 
\item All perturbation orders are left implicit. Here we keep them explicit as before.
\item The angles $(\theta,\phi)$ are written explicitly. This gives rise to some factors of $\sin\theta$ between expressions, which are implicit in our notation with $\theta^a$ and covariant derivatives.
\end{itemize}
The leading terms in the main result, Eq.~(94) are
\begin{equation}\label{e:YZ}
\Sigma^{(2)} = \delta^{(2)} + \delta V^{(2)} +  \delta^{(1)} \delta V^{(1)} \, .
\end{equation}
The first order volume perturbation is given by
\begin{align}
\delta V^{(1)} = -2\ka^{(1)} + H_z \partial_z \delta r^{(1)} = -2\ka^{(1)} + \HH^{-1} \partial_r^2 v^{(1)} \, ,
\end{align}
where we use $\dd_z = H_z^{-1}\dd_r$.
The product of these terms with $\delta^{(1)}$ is readily identified in Eq.~\eqref{e:reference}. Going on to the second order volume perturbation, we find
\begin{align}\label{e:dV2}
\delta V^{(2)} = \delta \mathrm D^{(2)}+ H_z \dd_z \delta r^{(2)} -2 H_z \ka^{(1)} \dd_z \delta r^{(1)} + \Delta x^{(1)b}\dd_b\delta V^{(1)} \, ,
\end{align}
where $\delta \mathrm D^{(2)}$ is the equivalent of $[|A|]^{(2)}$, which we show now. The dominating terms are
\begin{align}\label{e:DisA}
\delta \mathrm D^{(2)} =\frac{\partial}{\partial\theta} \de\theta^{(2)}+\frac{\partial}{\partial\phi}\de\phi^{(2)}
+\frac{\partial}{\partial\theta}\de\theta^{(1)}\frac{\partial}{\partial\phi}\de\phi^{(1)}
-\frac{\partial}{\partial\theta}\de\phi^{(1)}\frac{\partial}{\partial\phi}\de\theta^{(1)} \, .
\end{align}
Note the similarity to Eq.~(2.20) of Ref.~\cite{DiDio:2015bua}. Now we need to make sure the terms above are correctly calculated.  Concerning the angles, we will just look at $\theta$ and argue that with the appropriate factors $\sin\theta$ the calculations extend naturally to $\phi$. This also avoids confusion between the angle $\phi$ and the lensing potential. In the following, $\phi$ refers always to the potential, not the angle.  Let us first take the first order angles. Identifying $\mathcal A-\mathcal C_{\alpha\beta}e^\alpha e^\beta = 2\Psi$, they are given by
\begin{equation}
\delta\theta^{(1)} =- \int_0^{\bar r_z}\left(\frac{\bar r_z-\bar r}{\bar r_z\bar r}\right) \dd_\theta \left(\mathcal A-\mathcal C_{\alpha\beta}e^\alpha e^\beta\right)  dr = \dd_\theta\phi^{(1)},
\end{equation}
just as expected.  This shows that the last two terms of (\ref{e:DisA}) are simply $(\ka^{(1)})^2 -|\ga^{(1)}|^2$ of Eq.~(\ref{e:lensref}).  We want to show that the first two terms are equal to our $\nabla_a[\theta^a]^{(2)}$. For the second order angles, the leading parts are
\begin{equation}
\de\theta^{(2)}\bar r_z = - \int_0^{\bar r_z}\left(\bar r_z-\bar r\right)e_{\theta \alpha} \left(\delta\Gamma^{(2)\alpha}+\Delta x^{(1)b}\dd_b \delta\Gamma^{\alpha(1)} \right)  d\bar r \, .
\end{equation}
Here $e_{\theta \alpha}$ is the $\al$ component of the unit vector in $\theta$-direction. We expand the contracted Christoffel symbols, $ \delta\Gamma$, and interpret the sum in $\Delta x^{(1)b}\dd_b \delta\Gamma^{\alpha(1)}$ as a sum only over the two transversal directions. Substituting $\Delta x^{(1)b}\dd_b = \theta^{(1)a}\nabla_a$, the correct expression becomes\footnote{There is also a typo in eqs. (51) and (52) where the angular derivative in the third line should only act on the metric components and not on $\De x^b$.}
\begin{equation}
\de\theta^{(2)} = - \int_0^{\bar r_z}\left(\frac{\bar r_z-\bar r}{\bar r_z \bar r}\right)  \left(\dd^\theta\Psi^{(2)}+\theta^{(1)a}\nabla_a \dd^\theta \Psi^{(1)} \right)  d\bar r\,.
\end{equation}
This means the first two terms of Eq.~\eqref{e:DisA} can be written as
\begin{equation}
\nabla_a \theta^{(2)a} = \Delta_\Omega \phi^{(2)} -  \int_0^{\bar r_z}\left(\frac{\bar r_z-\bar r}{\bar r_z \bar r}\right) \nabla_b\left(\nabla^a\phi^{(1)}\nabla_a \nabla^b \Psi^{(1)}\right)  d\bar r \, .
\end{equation}
This formula matches Eq.~\eqref{e:angle2}, showing that indeed $\de\mathrm D^{(2)} = [|A|]^{(2)}$, all pure lensing terms are included. For the remaining terms of Eq.~\eqref{e:dV2}, we find the leading order contributions to be $\delta r^{(2)} = \HH^{-1}\partial_r v^{(2)} - \HH^{-1} \Delta x^b\partial_b \delta z^{(1)}$. Writing out the terms, we obtain
\begin{align}
 H_z \dd_z \delta r^{(2)} -2 H_z \kappa \dd_z \delta r^{(1)} =& \\
  \HH^{-1}\partial_r^2 v^{(2)} +& \HH^{-1}\nabla^a\phi^{(1)}\nabla_a \partial_r^2 v^{(1)} + \HH^{-2} \partial_r\big[ \partial_r v^{(1)} \partial_r^2 v^{(1)} \big] -2\HH^{-1}\ka^{(1)}\partial_r^2 v^{(1)} \, , \nonumber
\end{align}
which directly match the corresponding terms in Eq.~\eqref{e:reference}. Since $\delta^{(2)}$ is left untouched here, we are missing terms relating to the Taylor expansion of $\delta^{(1)}$, both radial and angular. Including these terms, we account for all leading order contributions, but are still left with the last term of Eq.~\eqref{e:dV2}, $\Delta x^{(1)b}\dd_b\delta V^{(1)}$. Expanding this expression gives
\begin{equation}
\Delta x^{(1)b}\dd_b\delta V^{(1)} = -2\nabla^a\phi^{(1)}\nabla_a\kappa^{(1)} + \HH^{-1}\nabla^a\phi^{(1)}\nabla_a \partial_r^2 v^{(1)} + \HH^{-2} \partial_r v^{(1)} \partial_r^3 v^{(1)} 
\,.
\end{equation}
 But   all these terms are already accounted for above in $\de\mathrm D^{(2)}$ and in $ H_z \dd_z \delta r^{(2)}$ respectively. The reason for this is clear. The three terms are the Taylor expansion of RSD and the lensing term which are contained in the expressions for $\de r^{(2)}$ and $\de\theta^{(2)}$ above. We argue that adding also $\Delta x^{(1)b}\dd_b\delta V^{(1)}$ is double-counting this effect\footnote{Jaiyul Yoo agrees with this finding (private communications).}. Discarding this last term entirely in (\ref{e:dV2}), the results (\ref{e:YZ}) and (\ref{e:reference}) agree.

\subsection{Discussion}
As already pointed out in~\cite{DiDio:2015bua}, the pure density and velocity terms of  (\ref{e:reference}) perfectly agree with Newtonian perturbation theory as presented e.g. in~\cite{Bernardeau:2001qr}.  The contribution to the bispectrum, 
$$\langle\De^{(2)}(\bn_1,z_1)\De^{(1)}(\bn_2,z_2)\De^{(1)}(\bn_3,z_3)\rangle + \text{cyclic} \;, $$
 from the dominant terms discussed here
has been computed numerically in~\cite{DiDio:2015bua}. There, it is found that for equal redshifts, $z_1=z_2=z_3$, the result is entirely dominated by the standard Newtonian terms. However, for  different redshifts, the Newtonian terms rapidly decay and the result is soon dominated by the new lensing contributions. For two equal and one different redshifts, the mixed lensing-standard terms dominate while for three different redshifts the pure lensing terms dominate the bispectrum.  The amplitude of the bispectrum at widely separated redshifts coming mainly from lensing amounts to about 10\% of the amplitude at equal redshifts, coming mainly from the standard Newtonian terms. The same behavior has  been found previously for the power spectrum~\cite{Montanari:2015rga}. Even though the lensing contribution is subdominant at equal redshifts $z_1=z_2$,  it dominates the signal for significantly different redshifts, $z_1\neq z_2$. This is quite intuitive as foreground galaxies cause the lensing potential and contribute to the integrated signal lensing signal from higher redshifts. We expect this behavior to remain true also for higher order correlations and for higher order  corrections to the power spectrum.

\begin{figure}[htb]
\begin{center}
\includegraphics[width=.6\textwidth]{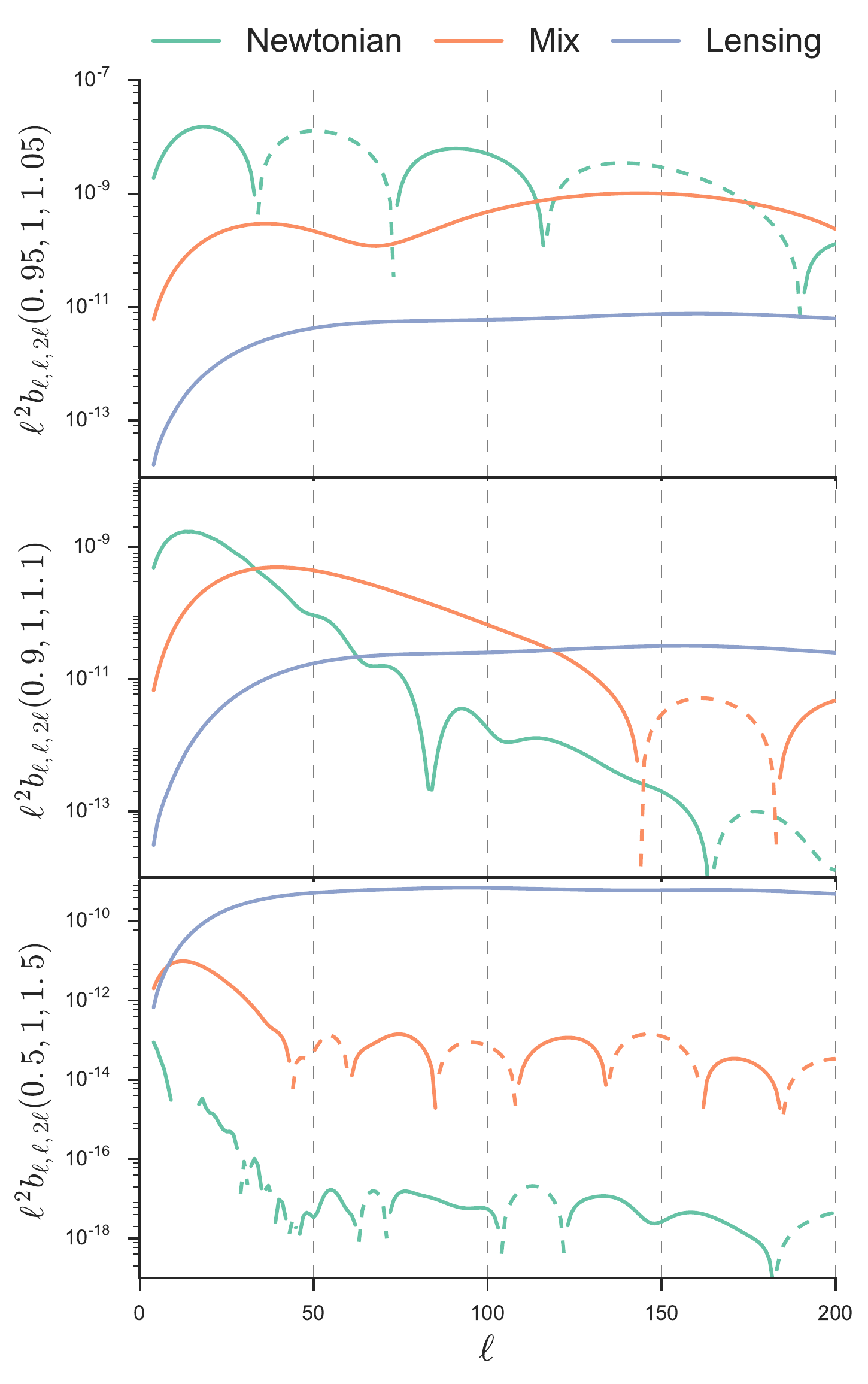}
\caption{The reduced bispectrum $\ell^2b_{\ell,\ell,2\ell}(z_1,z_2,z_3)$ is shown for very similar redshifts, $(z_1,z_2,z_3)=(0.95,1,1.05)$, somewhat different redshifts, $(z_1,z_2,z_3)=(0.9,1,1.1)$ and very different redshifts, $(z_1,z_2,z_3)=(0.5,1,1.5)$. The Newtonian terms (green) are compared with the lensing terms (blue) and their cross-correlations (orange). For small redshift separations the Newtonian terms dominate while for large redshift separation the lensing contributions dominate, see~~\cite{DiDio:2015bua} for details. }
\label{fig:bispectrum}
\end{center}
\end{figure}

The dominant projection effects are the redshift space distortions which contribute up to 30\% of the total signal at equal redshifts, both in the power spectrum and in the bispectrum, and the lensing terms which dominate the signal (99\% of the total) for widely separated redshifts.  Plots of both the spectrum and the bispectrum can be found in~\cite{Montanari:2015rga} and~\cite{DiDio:2015bua}.  For illustration we re-plot in Fig.~\ref{fig:bispectrum} the analog of Fig.~7 of Ref.~\cite{DiDio:2015bua}. The 'Newtonian' terms there are all terms containing only density, $\de$ and velocity $v$ and their rerivatives. These terms appear also in a purely Newtonian treatment as e.g. in Ref.~\cite{Bernardeau:2001qr}.
The 'lensing terms' are those containing $\kappa$, $\Psi_1$ or $\nabla_a\phi$.

\section{Extension to higher perturbative orders}\label{sec:higherorder}
The  procedure developed in the previous section is easily generalized to derive the expression for the dominant terms of higher perturbative orders in the number counts. Higher orders will in particular be important to calculate the 1-loop corrections to the power spectrum, where at least third order perturbative results  are needed.

Following the same steps, only this time to third order, we arrive at the following expression for the dominant third order terms,
\begin{align}
\Sigma^{(3)} =& \
[\delta]^{(3)}+[\text{\textsc{rsd}}]^{(3)} + [|{A^a}_b|]^{(3)} 
+ \delta^{(1)}[\text{\textsc{rsd}}]^{(2)} + [\delta^{(2)}]\text{\textsc{rsd}}^{(1)}
+\delta^{(1)}[ |{A^a}_b|]^{(2)} +[\delta^{(2)} ]|{A^a}_b|^{(1)} \nonumber\\
+&\ \text{\textsc{rsd}}^{(1)}[ |{A^a}_b|]^{(2)}  +[\text{\textsc{rsd}}]^{(2)} |{A^a}_b|^{(1)}
+ \delta^{(1)}\text{\textsc{rsd}}^{(1)}  |{A^a}_b|^{(1)} \, , \label{e:3rdorder}
\end{align}
where all first and second order terms have already been calculated. The brackets around the second and third order terms serve as a reminder that these are not simply the corresponding order term at the unperturbed position but we also have to take into account the lower order terms $X^{(3-n)}$ with the deviation from the Born approximation at order $n$.  Note also that the separation of second order terms is important, as it determines which cross terms appear at third order. The third order determinant of the lensing map can be written analogously to (\ref{e:lensref})
\begin{align}
[|{A^a}_b|]^{(3)} = \nabla_a[\theta^a]^{(3)} + \nabla_a\theta^{(1)a}\nabla_b[\theta^b]^{(2)} - \nabla_a\theta^{(1)b}\nabla_b[\theta^a]^{(2)},
\end{align}
and the third order analogue of Eq.~\eqref{e:2ndexp} is given by
\begin{align}
[F( r(z+\de z)(\bn & +\de\bn) )]^{(3)} =  \\
\phantom{=}&\ F^{(3)}+\HH^{-1} \dd_r v^{(1)} \dd_{r} F^{(2)} + \nabla^a\phi^{(1)}\nabla_a F^{(2)} \nonumber\\
+&\ \HH^{-1} \left[ \dd_r v^{(2)} + \HH^{-1} \dd_r v^{(1)} \dd_r^2 v^{(1)} + \nabla^a\phi^{(1)}\nabla_a \dd_r v^{(1)} \right] \dd_{r} F^{(1)}  \nonumber\\
+&\ \left[ \nabla^a\phi^{(2)} -2 \int_0^{r(z)}dr\frac{r(z)-r}{r(z)r} \nabla^b\phi^{(1)}\nabla_b \nabla^a\Psi^{(1)} \right]
\nabla_a F^{(1)}\nonumber \\
+&\ \frac{1}{2} \left[ \HH^{-2} (\dd_r v^{(1)})^2 \dd_{r}^2 F^{(1)}  + \nabla^a\phi^{(1)}\nabla^b\phi^{(1)}\nabla_b\nabla_a F^{(1)}\right] \nonumber \\
+&\ \HH^{-1}\dd_r v^{(1)} \nabla^a\phi^{(1)} \dd_r\nabla_a F^{(1)}  \, . \nonumber
\end{align}
On the right hand side all quantities are now evaluated at the unperturbed positions.
Here we have inserted the higher order 'bare' radial and transversal shifts,
\begin{align}
 r^{(n)} =H^{-1}  z^{(n)} &= \HH^{-1}\dd_r v^{(n)} \nonumber \\
\theta^{(n)a} &= -2 \int_0^{r(z)}dr\frac{r(z)-r}{r(z)r} \nabla^a \Psi^{(n)} \equiv \nabla^a\phi^{(n)} \ .\nonumber
\end{align}
The third order quantities $\de^{(3)}, v^{(3)}, \Psi^{(3)}$ are given in App.~\ref{app:Newton}.

The third order leading contribution to the angles has already been calculated in Ref.~\cite{Fanizza:2015swa} and we check our simple prescription against their result. We obtain to third order, by including second order deviations of the photon path,
\begin{align}
[\theta^a]^{(3)} =& -2\!\int_0^{r(z)}\hspace{-1mm}dr\frac{r(z)-r}{r(z)r} \Big( \nabla^a\Psi^{(3)} \!+ \!\nabla^b\phi^{(1)}\nabla_b\nabla^a\Psi^{(2)}+[\theta^{b}]^{(2)}\nabla_b\nabla^a\Psi^{(1)} + \nonumber \\ 
 & \hspace{4.1cm} \frac{1}{2}\nabla^b\phi^{(1)}\nabla^c\phi^{(1)}\nabla_b\nabla_c\nabla^a\Psi^{(1)}\Big) \nonumber \\
=&\ \nabla^a\phi^{(3)} -2\int_0^{r(z)}dr\frac{r(z)-r}{r(z)r}
\Bigg(\nabla^b\phi^{(1)}\nabla_b\nabla^a\Psi^{(2)} + \nabla^b\phi^{(2)}\nabla_b\nabla^a\Psi^{(1)} \nonumber\\
- &\ 2\left[ \int_0^{r}dr'\frac{r-r'}{r \ r'} \nabla^c\phi^{(1)}\nabla_c \nabla^b\Psi^{(1)} \right] \nabla_b\nabla^a\Psi^{(1)}
+ \frac{1}{2}\nabla^b\phi^{(1)}\nabla^c\phi^{(1)}\nabla_b\nabla_c\nabla^a\Psi^{(1)} \Bigg) \, .
\end{align}
The first two lines here match exactly the dominant contribution in the result of \cite{Fanizza:2015swa}, when taking into account factors $\frac{1}{2}$ and $\frac{1}{3!}$ in their definitions of second and third order metric perturbations.

\section{Second order effects on the power spectrum: a preliminary study}\label{s:num}
We now briefly discuss the effect of higher order perturbations on the correlation function and the power spectrum, i.e. the $C_\ell$s. The 1-loop correction to the correlation function is
\begin{align}
\xi^{(2)}(\bn\cdot\bn',z,z') &=  \langle \Si^{(2)}(\bn,z) \Si^{(2)}(\bn',z')\rangle - \langle \Si^{(2)}(\bn,z)\rangle\langle \Si^{(2)}(\bn',z')\rangle \nonumber \\ 
&+ \langle\Si^{(1)}(\bn,z) \Si^{(3)}(\bn',z') \rangle + \langle\Si^{(3)}(\bn,z) \Si^{(1)}(\bn',z') \rangle \nonumber \\
&\equiv \frac{1}{4\pi} \sum_\ell (2\ell+1)C_\ell^{(2)}(z,z')  P_\ell (\bn\cdot\bn') \label{e:2ndxi}
\end{align}
These corrections converge very badly (maybe not at all) similar to higher order corrections e.g. to the luminosity distance~\cite{Clarkson:2011uk}.
Furthermore, like for the power spectrum in Fourier space, see e.g.~\cite{Baumann:2010tm,Pajer:2013jj}, there are 'counter terms' which have to be added to this naive expression in order to obtain a consistent result. We leave a rigorous study of this as a future project.

Here we just derive the formal expressions for the simplest terms in the 1-loop power spectrum in $\ell$-space. First we show how second order quantities from squared first order terms contribute. Denoting the  different first order terms by $\De^A,\De^B,\De^C,\De^D$ (eg. $\de^{(1)},\dd_r^2 v^{(1)}$ etc.), we consider the following second order contributions (which is \emph{not} completely general),
\begin{align}
\De^{AB}(\bn,z) \equiv (\De^A\cdot\De^B)(\bn,z)  \ .
\end{align}
The contribution from such product terms to the 1-loop correlation function is
\begin{align}\label{e:2ndindfacs}
&\xi^{AB|CD}(\bn\cdot\bn',z,z') \equiv \langle\De^{AB}(\bn,z)\De^{CD}(\bn',z')\rangle - \langle\De^{AB}(\bn,z)\rangle\langle\De^{CD}(\bn',z')\rangle \nonumber\\
=&\langle \De^{A}(\bn,z) \De^{C}(\bn',z')\rangle\langle \De^{B}(\bn,z) \De^{D}(\bn',z')\rangle +
\langle \De^{A}(\bn,z) \De^{D}(\bn',z')\rangle\langle \De^{B}(\bn,z) \De^{C}(\bn',z')\rangle\nonumber\\
=&  \xi^{AC}(\bn\cdot\bn',z,z') \xi^{BD}(\bn\cdot\bn',z,z') +  \xi^{AD}(\bn\cdot\bn',z,z') \xi^{BC}(\bn\cdot\bn',z,z') \, . 
\end{align}
These are simply products of first order correlation functions of the factors. We can use this to compute the corresponding contribution to the power spectrum. We first write out the first order correlation functions in terms of the $C_\ell(z,z')$ which can be calculated with the help of eg. \textsc{class}, \cite{Blas:2011rf,DiDio:2013bqa},
\begin{align}
\xi^{AB}&= \frac{1}{4\pi} \sum_\ell (2\ell+1)C_\ell^{AB}(z,z')  P_\ell (\bn\cdot\bn') \Rightarrow\nonumber\\
\xi^{AB|CD}(\bn\cdot\bn',z,z') &= \frac{1}{(4\pi)^2}\sum_{\ell,\ell'}(2\ell+1)(2\ell'+1) \\
&\times \big[ C^{AC}_\ell(z,z') C^{BD}_{\ell'}(z,z') + C^{AD}_\ell(z,z') C^{BC}_{\ell'}(z,z')  \big]
P_\ell(\bn\cdot\bn')P_{\ell'}(\bn\cdot\bn') \,,  \nonumber
\end{align}
where $P_\ell$ denotes the Legendre polynomial of order $\ell$. We would now like this in the form of Eq.~\eqref{e:2ndxi}. To do this, we use the following expansion of products of Legendre polynomials,
\begin{align}
P_\ell(x)P_{\ell'}(x) = 
\sum_{L=|\ell-\ell'|}^{\ell+\ell'} 
\thJ{\ell}{\ell'}{L}{0}{0}{0}^2 (2L+1)P_L(x) \, ,
\end{align}
which is a special case of the expansion of a product of spherical harmonics.  
The squared Wigner 3j symbols can in this case be written explicitly as
$$
\thJ{\ell}{\ell'}{L}{0}{0}{0}^2\!\!= \frac{[(\ell-\ell'+L-1]!![(\ell'-\ell+L-1]!![(\ell+\ell'-L-1]!![(\ell+\ell'+L)/2]! }{[(\ell-\ell'+L)/2]![(\ell'-\ell+L)/2]![(\ell+\ell'-L)/2]! [(\ell+\ell'+L-1]!!(\ell + \ell' + L + 1)} \, ,
$$
when $\ell+\ell'+L$ is even and the triangle inequality is satisfied, and is 0 otherwise.
This means the contribution from these terms to 1-loop power spectrum is
\begin{align}
C^{AB|CD}_\ell(z,z') = \sum_{\ell_1\ell_2} &\frac{(2\ell_1+1)(2\ell_2+1)}{4\pi}\thJ{\ell}{\ell_1}{\ell_2}{0}{0}{0}^2 \nonumber \\
&\times\left(C_{\ell_1}^{AC}(z,z') C_{\ell_2}^{BD}(z,z') +C_{\ell_1}^{AD}(z,z') C_{\ell_2}^{BC}(z,z') \right)
   \, . \label{e:cl22}
\end{align}

Pure product contributions from third order are structurally even simpler.  We set $\De^{ABD}=\De^A\De^B\De^C$. Since three of the four factors are evaluated at the same position and redshift, we get
\begin{align}
\xi^{ABC|D}(\bn\cdot\bn',z,z') &\equiv \langle \De^{ABC}(\bn,z) \De^{D}(\bn',z')\rangle \nonumber\\
&=\xi^{BC}(1,z,z)\xi^{AD}(\bn\cdot\bn',z,z') \nonumber \\
&+ \xi^{AC}(1,z,z)\xi^{BD}(\bn\cdot\bn',z,z') \\
&+\xi^{AB}(1,z,z)\xi^{CD}(\bn\cdot\bn',z,z') \nonumber \, ,
\end{align}
where half of the functions are evaluated at $\bn\cdot\bn=1$. Since $P_\ell(1)=1$, these are simply given by
\begin{equation}
\xi^{AB}(1,z,z) = \frac{1}{4\pi}\sum_\ell (2\ell+1) C^{AB}_\ell(z) \, ,
\end{equation}
and we can write this contribution to the 1-loop power spectrum as
\begin{align}
C_\ell^{ABC|D}(z,z') =& \frac{1}{4\pi} \sum_{\ell'} (2\ell'+1) \nonumber \\
\times&\left\{
C_{\ell'}^{BC}(z) C_\ell^{AD}(z,z')
+C_{\ell'}^{AC}(z)C_\ell^{BD}(z,z')
+ C_{\ell'}^{AB}(z) C_\ell^{CD}(z,z')
\right\} \, . \label{e:cl31}
\end{align}

Of course these corrections to the power spectrum are mainly relevant in the weakly non-linear regime and perturbation theory is not expected to converge in the fully non-linear regime. Nevertheless, perturbation theory remains computationally much less heavy than N-body simulations and  represents an important cross-check of the latter, especially for simulations which attempt ray-tracing.

However, many of the above sums over $\ell$ actually seems to diverge and for a correct calculation of the corrections to the power spectrum, we have to introduce counter terms. A similar problem is encountered when calculating the  luminosity distance at second order~\cite{Clarkson:2011uk}. But also Newtonian perturbation theory in Fourier space at second order requires the introduction of phenomenological counter terms~\cite{Pajer:2013jj,Senatore:2014vja}. The  determination and analysis of these counter terms, which will  differ in the directly observable $\ell$-space from those in Fourier space, is an interesting  future project but it is beyond the scope of the present work. We do expect new counter terms from the lensing contributions. Furthermore, due to the fact that we calculate a directly observable quantity, infrared divergences, present in the density perturbation spectrum~\cite{Noh:2009yu} should should not be present in $\ell$-space.

\section{Conclusions}\label{s:con}
We have re-derived the relativistic second order contribution to the cosmological galaxy number counts.  We have concentrated on the terms which are dominant in powers of $k/\HH$ i.e. of order $(k/\HH)^4\Psi^2$. The disagreement between previous results in the literature are identified and clarified. With a clear physical picture in mind, we  generalize the second order expression to recursively obtain the leading number count contributions at any order in perturbation theory.  We explicitly write down the new terms occuring at third order. This allows, in principle,  to compute  the 1-loop corrections to the correlation function and to the power spectrum in observable $\ell$ and redshift space, by including second and third order contributions.

The dominant terms in the relativistic number counts come from density, redshift space distortions and lensing. The former two, the so called 'Newtonian terms' are in perfect agreement with the Newtonian analysis. The lensing terms are only present in a relativistic treatment. They are subdominant at equal redshifts but dominate the result for different redshifts. There are of course many more contributions to the second order number counts. They can be classified as terms of order $(k/\HH)^3\Psi^2$, $(k/\HH)^2\Psi^2$, 
$(k/\HH)\Psi^2$ and $\Psi^2$. The $(k/\HH)^2\Psi^2$-terms in the power spectrum of the first order number counts lead to corrections  which are degenerate with those coming from a primordial local non-Gaussianity, $f_{\rm NL}$, and they have to be calculated carefully in order not to mistake them for a non-Gaussianity~\cite{Camera:2014sba,Raccanelli:2015vla}. Recently, the contributions to the bispectrum from  the terms of the last group, $\propto \Psi^2$ have been estimated in the squeezed limit~\cite{Kehagias:2015tda}.  Clearly, the computation of  the observed number count bispectrum 
is still in its infancy, especially for the very large scales. In this paper, we have found the full expression for the first group of terms which dominate on scales well within the horizon. A detailed comparison of the remaining groups of terms published in Refs.~\cite{Bertacca:2014wga,Yoo:2014sfa,DiDio:2014lka} is still missing and some discrepancy may still remain.

\acknowledgments
RD thanks Daniele Bertacca, Giuseppe Fanizza, Roy Maartens and Giovanni Marozzi for discussions.  We are grateful to Filippo Vernizzi for clarifications and to Jaiyul Yoo for comments. JTN thanks the University of Geneva for hospitality. This work is supported by the Swiss National Science Foundation and by Danmarks Grundforskningsfond under grant no. 1041811001.

\appendix
\section{The higher order Newtonian density and velocity perturbations}\label{app:Newton}
Here we repeat the second and third order expressions for the Newtonian density and velocity perturbations in Fourier space. They are obtained from the expansion of the continuity, Euler and Poisson equations. They can be found e.g. in~\cite{Bernardeau:2001qr}. 

\bea\label{e:de2}
\de^{(2)}(\bk,t)&=&  \frac{1}{\left( 2 \pi \right)^3} \int d^3k_1  F_2 \left( \bk_1, \bk-\bk_1 \right) \delta \left( \bk_1 ,t\right) \delta\left( \bk-\bk_1,t \right)\, ,\\
v^{(2)}(\bk,t) &=& -\frac{\HH}{k^2}   \frac{f^2(z) }{\left( 2 \pi \right)^3} \int d^3k_1G_2 \left( \bk_1,  \bk-\bk_1 \right) \delta\left( \bk_1 ,t\right) \delta \left( \bk-\bk_1,t \right)\, ,\label{e:v2} \quad \\
\Psi^{(2)}(\bk,t) &=& -\frac{3\HH^2\Om_m(t)}{2k^2}  \de^{(2)}(\bk,t) \,. \label{e:ka2}
\eea
Here  $f(z)=d\log D_1/d\log a$ is the growth factor and $D_1$ is the linear growth rate of density perturbations. The kernels $F_2$ and $G_2$ are given by~\cite{Goroff:1986,Bernardeau:2001qr}
\bea
F_2(\bk_1,\bk_2) &=&\frac{5}{7} + \frac{1}{2} \frac{\bk_1 \cdot \bk_2}{k_1 k_2} \left( \frac{k_1}{k_2} + \frac{k_2}{k_1} \right) + \frac{2}{7} \left(\frac{\bk_1 \cdot \bk_2}{k_1 k_2} \right)^2 \nonumber \\   \label{e:F2}
 &=&\frac{17}{21} + \frac{1}{2} \left( \frac{k_1}{k_2} +  \frac{k_2}{k_1} \right) P_1 \left( \hat \bk_1 \cdot \hat \bk_2\right) + \frac{4}{21} P_2 \left( \hat \bk_1 \cdot \hat \bk_2 \right) \,,\\
 G_2(\bk_1,\bk_2) &=&\frac{3}{7} + \frac{1}{2} \frac{\bk_1 \cdot \bk_2}{k_1 k_2} \left( \frac{k_1}{k_2} + \frac{k_2}{k_1} \right) + \frac{4}{7} \left(\frac{\bk_1 \cdot \bk_2}{k_1 k_2} \right)^2  \nonumber \\
 &=&\frac{13}{21} + \frac{1}{2} \left( \frac{k_1}{k_2} +  \frac{k_2}{k_1} \right) P_1 \left( \hat \bk_1 \cdot \hat \bk_2\right) + \frac{8}{21} P_2 \left( \hat \bk_1 \cdot \hat \bk_2 \right) \,,\label{e:G2}
 \eea
where $P_1$ and $P_2$ denote the first and second order Legendre polynomials. For more details, see~\cite{Bernardeau:2001qr}. The third order expressions are similarly

\begin{align}\label{e:de3}
\de^{(3)}(\bk,t)&=  \frac{1}{\left( 2 \pi \right)^6} \int d^3k_1d^3k_2  F_3 \left( \bk_1, \bk_2,\bk-\bk_1 -\bk_2\right) \delta \left( \bk_1 ,t\right)\delta \left( \bk_2 ,t\right) \delta\left( \bk-\bk_1-\bk_2,t \right)\, , \\
v^{(3)}(\bk,t) &= -\frac{\HH}{k^2}   \frac{f^2(z) }{\left( 2 \pi \right)^6} \int d^3k_1G_3 \left( \bk_1, \bk_2,\bk-\bk_1 -\bk_2 \right) \delta\left( \bk_1 ,t\right) \delta\left( \bk_2 ,t\right) \delta \left( \bk-\bk_1-\bk_2,t \right)\, ,\label{e:v3} \\
\Psi^{(3)}(\bk,t) &= -\frac{3\HH^2\Om_m(t)}{2k^2}  \de^{(3)}(\bk,t) \,. \label{e:ka3}
\end{align}
The third order kernels can be written in terms of the second order kernels as \cite{Bernardeau:2001qr}
\begin{align}
F_3(\bk_1,\bk_2,\bk_3) &= \frac{1}{18}\big[G_2(\bk_1,\bk_2)[7\alpha(\bk_1+\bk_2,\bk_3)+4\beta(\bk_1+\bk_2,\bk_3)] \nonumber\\
&+7\alpha(\bk_1,\bk_2+\bk_3)F_2(\bk_2,\bk_3)
\big]\ , \\
G_3(\bk_1,\bk_2,\bk_3) &= \frac{1}{6}\big[G_2(\bk_1,\bk_2)[\alpha(\bk_1+\bk_2,\bk_3)+4\beta(\bk_1+\bk_2,\bk_3)] \nonumber\\
&+\alpha(\bk_1,\bk_2+\bk_3)F_2(\bk_2,\bk_3)
\big] \ ,
\end{align}
where the mode-coupling functions are given by
\begin{align}
\alpha(\bk,\bk') = \frac{(\bk+\bk')\cdot\bk}{k^2} \ , \\
\beta(\bk,\bk') = \frac{(\bk+\bk')^2\bk\cdot\bk'}{2k^2k'^2} \ .
\end{align}

\vspace{2cm}
\begingroup\raggedright\endgroup

\end{document}